\documentclass{vgtc}                        
\graphicspath{{figures/}{pictures/}{images/}{./}}
\usepackage{times}                 
     
\usepackage{tabu}             
\usepackage{booktabs}          
\usepackage{lipsum}            
\usepackage{mwe}               
\usepackage{mathptmx} 
\usepackage{enumitem}
\usepackage{xspace}
\usepackage{fullpage}
\onlineid{0}
\vgtccategory{Research}
\vgtcinsertpkg

\title{RaivenTracks: Branching Provenance for Conversational Visualization Workflows}

\author{Ella Hugie\thanks{Ella Hugie and Alexandra Irger are co-first authors.}\\ %
        \scriptsize Harvard University %
\and Alexandra Irger\footnotemark[1]\\ %
     \scriptsize Harvard University
\and Grace Guo\\ %
     \scriptsize Harvard University\\
\and Kenneth Moreland\\ %
     \scriptsize Oak Ridge National Laboratory\\
\and David Pugmire\\ %
     \scriptsize Oak Ridge National Laboratory\\
\and Scott Klasky\\ %
     \scriptsize Oak Ridge National Laboratory\\
\and Hanspeter Pfister\\ %
     \scriptsize Harvard University\\
}

\teaser{
  \centering
\includegraphics[width=\linewidth]{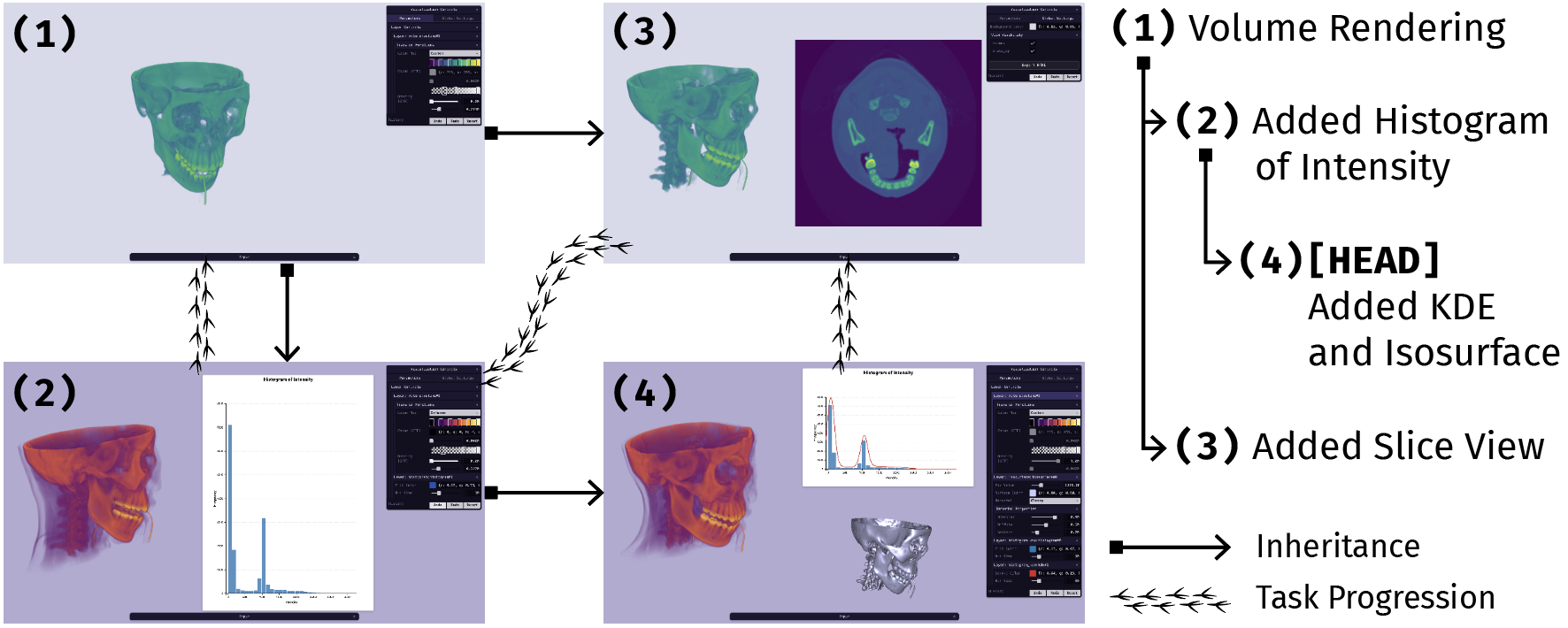}
  \caption{\newsysname~maintains persistent conversational state and a tree-based version history over validated RaivenDSL specifications, enabling users to revisit, branch from, and recover prior visualization states.
  This branchable history is a step toward provenance support for future scientist-in-the-loop AI-driven workflows.
  }
  \label{fig:teaser}
}

\newcommand{\sysname}{Raiven\xspace}
\newcommand{\newsysname}{RaivenTracks\xspace}

\abstract{

As AI agents increasingly participate in scientific workflows, scientists are shifting from direct authorship toward oversight, inspection, and steering. LLM-driven visualization systems are a promising interface for this hand-off, yet they remain largely stateless, forcing users to reconstruct context across refinements and offering little support for revisiting prior decisions or exploring alternatives. We present \textit{\newsysname}, a workflow-aware extension of the Raiven DSL-mediated visualization pipeline that treats validated visualization specifications as persistent, branchable checkpoints. Because each checkpoint is a verifiable RaivenDSL specification rather than a dialogue transcript, restoring a node recompiles a known artifact rather than re-interpreting prior context. \newsysname~contributes a two-level state management architecture that pairs a persistent, branchable version tree with a fine-grained undo/redo stack over runtime visualization settings, across both InfoVis and SciVis backends. A formative pilot study with three visualization researchers shows early promise, with all participants adopting the version tree for branching and recovery, and surfaces design directions for tree navigation, node labeling, and scalability that inform a planned controlled comparison against Raiven without version history.
We frame branchable conversational visualization history as a step toward provenance support for future scientist-in-the-loop oversight of AI-driven scientific workflows.
}


\begin{document}
\maketitle

\section{Introduction}

AI agents are increasingly being integrated into scientific workflows, where they can help generate hypotheses, design experiments, and refine analyses with limited human intervention~\cite{king2009robot,stach2024autonomous}. As this happens, scientists do not disappear from the loop; their role shifts toward inspecting machine-generated artifacts, recovering from errors or unproductive directions, and deciding when to redirect an evolving workflow. Visual analytics can support this hand-off by making intermediate states visible and actionable, but doing so requires provenance mechanisms that expose how analytic states evolve over time. Without them, scientists may see only a workflow's latest output, with limited ability to understand how it was produced or to return to earlier decision points.

LLM-driven visualization assistants make it increasingly practical to generate and refine visualization specifications from natural-language prompts, lowering the barrier to authoring~\cite{dibia2023lida,tian2024chartgpt,maddigan2023chat2vis}. However, visualization is inherently iterative: researchers revise encodings, remap variables, correct intermediate decisions, and explore alternative representations as hypotheses evolve. This pattern is well documented, as sensemaking and exploratory workflows proceed through cycles of hypothesis generation, visual inspection, and progressive refinement rather than single-shot specification~\cite{heer2008graphicalhistories,scheidegger2013vistrails, lin2020dziban}.

Despite rapid progress in conversational text-to-visualization, many systems and benchmarks still emphasize one-step generation from user prompts, leaving workflow-level refinement comparatively under-supported~\cite{luo2021nvbench,dibia2023lida,maddigan2023chat2vis,tian2024chartgpt}. Prior natural-language interfaces demonstrate the importance of iterative clarification~\cite{gao2015datatone,setlur2016eviza,kumar2016articulate}, yet they treat refinements as re-specifications or implicit context updates. When scientists need to recover from a misinterpreted output, return to an earlier decision, or compare competing alternatives, current conversational visualization systems offer no principled way to inspect past states, create parallel refinement paths, or understand how a visualization evolved: operations that provenance and versioning research has shown to be central to analytic work~\cite{heer2008graphicalhistories,scheidegger2013vistrails,cutler2020trrack}.

\sysname~introduced a verifiable, DSL-mediated pipeline that separates interpretation from execution (Section~\ref{sec:background})~\cite{raiven}. Its DSL representation is a natural anchor for persistent state: rather than re-generating specifications from scratch, follow-up requests compile as structured updates to a well-defined artifact. Because that artifact is a verifiable specification rather than a dialogue transcript, it makes the specification, not the dialogue, the natural unit of versioning. \sysname, however, has no mechanism for maintaining this state across turns or exposing it as a navigable history.

We present \newsysname, a workflow-aware extension that adds versioning to DSL-based conversational visualization. \newsysname~maintains visualization specifications as a branchable version tree, letting users restore earlier states, branch from prior conversational contexts, and navigate divergent refinement paths across InfoVis and SciVis backends without manually reconstructing context.
%We view this branchable DSL history as a step toward future scientist-in-the-loop oversight of AI-driven scientific workflows, and treat that oversight setting as a motivating vision rather than a claim this study evaluates.
We view this branchable DSL history as a step toward scientist-in-the-loop oversight, and treat that oversight setting as a motivating vision rather than a claim this study evaluates.

Our contributions are: (1)~a two-level state management architecture pairing a persistent, branchable version tree over verifiable DSL checkpoints with a stack-based undo/redo layer for runtime visualization settings;
%designed so that restoring a node recompiles a known specification rather than re-interpreting dialogue;
(2)~a formative pilot study with three visualization researchers, including post-session interviews, that offers early evidence of version-tree adoption and informs the design of branchable conversational history, ahead of a planned controlled comparison against Raiven without version history; and (3)~design directions, grounded in the pilot, for scaling branchable conversational visualization histories toward provenance and scientist-in-the-loop oversight.
%in future AI-driven workflows.

\section{Related Work}
\label{sec:related}

\newsysname~sits at the intersection of three active research areas that inform the design of stateful, provenance-aware conversational visualization: LLM-driven visualization generation and iterative refinement~(\ref{sec:llm4vis}), DSLs for verifiable visualization specification~(\ref{sec:dsl4vis}), and provenance tracking~(\ref{sec:provenance}). We discuss each briefly and identify the shared gap that \newsysname~addresses.

\subsection{LLM-Driven Visualization Generation and Iterative Refinement}
\label{sec:llm4vis}
Recent surveys document rapid growth in LLM-assisted visualization authoring, exploration, and explanation, while noting open challenges in contextual grounding, multi-step reasoning, and end-to-end evaluation~\cite{brossier2026state}. Authoring systems such as LIDA~\cite{dibia2023lida}, ChartGPT~\cite{tian2024chartgpt}, and Chat2VIS~\cite{maddigan2023chat2vis} demonstrate that natural-language prompting can lower the barrier to visualization creation. LLM-driven interaction is also emerging in scientific visualization: Ai et al.\ introduce a natural-language interface for volumetric scene exploration via multi-agent LLM orchestration~\cite{ai2025nli4volvis}, where misinterpretation carries real analytic cost, though it manages context within an active session rather than exposing it as a persistent history to branch from or restore.
%More generally, most systems treat state implicitly as dialogue history and regenerate specifications each turn rather than maintaining a structured, validatable representation, leaving scientists no stable reference point from which to inspect or restore prior decisions.

Refinement is equally central. NL2VIS and mixed-initiative systems have long shown that users refine visualizations through iterative clarification and follow-up queries grounded in the current view~\cite{gao2015datatone,setlur2016eviza,kumar2016articulate, yu2019flowsense}. Mitra et al. show that many NL toolkits primarily support one-off utterances, and propose specification-level augmentations for multi-turn ambiguity resolution~\cite{mitra2022facilitating}, while benchmarks such as nvBench~\cite{luo2021nvbench} and CoVis~\cite{song2023covis} underscore the need for context modeling across turns.
Yet across these systems, refinements are treated as re-interpretations of the latest state or implicit updates to dialogue history, rather than updates to an explicit, validatable specification.
This is manageable in general-purpose tools, but becomes critical in scientific workflows, where recovering from a misinterpreted output or comparing competing analyses carries real analytic cost.

\subsection{DSLs for Verifiable Visualization Specification}
\label{sec:dsl4vis}
Declarative visualization DSLs, motivated by the Grammar of Graphics~\cite{wilkinson2005grammar} and operationalized in Vega-Lite~\cite{satyanarayan2017vegalite} and Draco~\cite{moritz2019draco}, enable deterministic compilation and retargeting across rendering backends. Heer and Bostock argue for separating specification from execution to simplify authoring and enable systematic reasoning about interaction~\cite{heer2010declarative}, and McNutt's survey motivates DSLs as a practical interface layer for both human authors and computational agents, since structured representations facilitate validation and incremental modification~\cite{mcnutt2022no}. DSL development is more mature on the InfoVis side than in SciVis. On the SciVis side, ViSlang~\cite{rautek2014vislang} composes procedural and declarative sublanguages for GPU-based visualization, and Shih et al.~\cite{shih2019declarative} introduce a declarative JSON grammar for volume visualization pipelines, but neither targets the full range of InfoVis mark types, and neither is designed as a generation target for language models. FlowNL~\cite{huang2022flownl} maps natural language to a declarative dataflow specification, but only within a single domain. \sysname~\cite{raiven} provides a verifiable DSL spanning both InfoVis and SciVis (Section~\ref{sec:background}). Verifiability matters beyond engineering convenience: it lets scientists inspect, interpret, and intervene in the specification process, making the DSL an essential tool for scientific oversight rather than merely an implementation detail.

%\subsection{Conversational Interfaces and Iterative Refinement}

\subsection{Provenance and Versioning in Visualization Workflows}
\label{sec:provenance}
Iterative analytic workflows benefit from structured support for revision, recovery, and comparison across alternatives~\cite{heer2008graphicalhistories,scheidegger2013vistrails,cutler2020trrack,xu2020survey}. Graphical Histories maps the design space of interactive history mechanisms, showing how branching timelines support analysis and replication~\cite{heer2008graphicalhistories}. VisTrails records change-based provenance for computational workflows, capturing the actions that transform one dataflow pipeline into another in a version tree, so any prior workflow can be selected, inspected, and re-executed~\cite{scheidegger2013vistrails}. Production tools offer related support, for example ParaView's session-state save/restore and action tracing, though its native history is linear rather than branching~\cite{ahrens2005paraview}. Trrack captures provenance in web-based visualizations, recording application state and interaction history independent of any specification language or domain~\cite{cutler2020trrack}.

More recently, Urbanite~\cite{moreira2025urbanite} couples an LLM-driven dataflow with a branchable provenance tree over dataflow versions for urban visual analytics. Its versioned unit is a dataflow whose nodes mix declarative visualization grammars with executable Python; \newsysname~instead versions each state as a single validatable RaivenDSL specification, paired with a fine-grained undo/redo history over runtime settings.
These systems assume the versioned artifact is something a human builds directly, a pipeline or a sequence of interactions, so that replaying the recorded history deterministically reproduces the state. Conversational LLM authoring breaks that assumption: there is no hand-built pipeline, the primary trace is a dialogue, and the artifact is generated rather than authored. Versioning the dialogue would not help, since replaying it re-invokes the model non-deterministically and surfaces the conversation rather than the verified state a scientist needs to inspect. \newsysname~recovers deterministic, VisTrails-style branching provenance precisely where that transfer fails, by grounding the version tree in the one artifact a conversational pipeline produces that is both verifiable and recompilable, the RaivenDSL specification. This grounding also gives a per-step record of what an agent produced, a substrate that work on LLM hallucination detection and recovery~\cite{huang2025hallucination} could act on if provenance is to support oversight of agent decisions, not only human ones.

These threads otherwise remain largely separate: DSLs supply a checkable specification, conversational interfaces supply iterative refinement, and provenance systems supply branching history, but none grounds a branchable history in the single verifiable specification a conversational pipeline produces each turn.
\newsysname~occupies this intersection, coupling the verifiable pipeline of \sysname~\cite{raiven} with persistent conversational state and a tree-structured version history over DSL checkpoints.

\begin{figure}[h!]
\centering
\includegraphics[width=\linewidth]{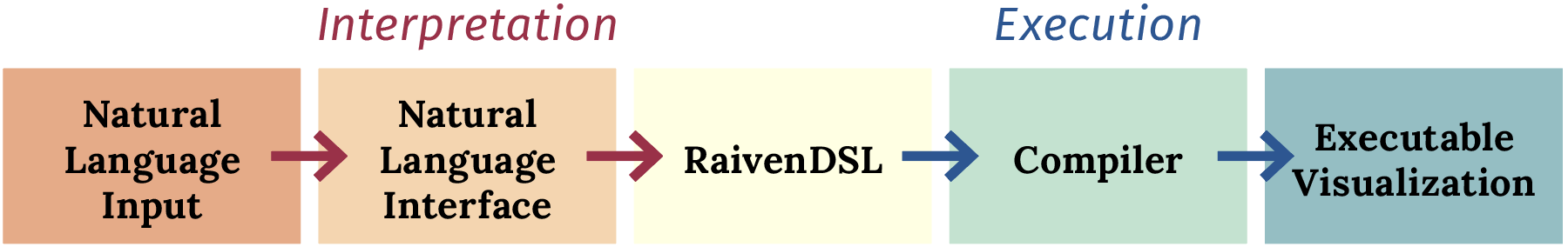}
\caption{\textbf{The Raiven~\cite{raiven} pipeline.} \emph{Interpretation} translates a natural-language request into a RaivenDSL specification via schema-guided LLM prompting; \emph{execution} compiles that specification deterministically to D3/VTK.js. \newsysname~attaches persistent, branchable state to the intermediate RaivenDSL artifact, the verifiable hand-off point between the two phases.}
\label{fig:raivenpipeline}
\end{figure}

\section{Background on Raiven}
\label{sec:background}
\newsysname~builds on \sysname~\cite{raiven}, a conversational visualization system that mediates natural-language authoring through a verifiable DSL rather than generating backend code directly (Figure~\ref{fig:raivenpipeline}). \sysname~separates \emph{interpretation} from \emph{execution}.
During interpretation, the LLM never emits rendering code: a sequence of narrowly scoped, individually validated prompts populates a structured session schema of typed data sources and a view structure of layers, marks, encodings, and linked selections. A final prompt translates it into RaivenDSL.
Because the model sees only dataset metadata and never raw values, generation is independent of context-window limits and cannot silently fabricate data.

The resulting specification is handed to a
deterministic compiler
that validates it independently of any backend, assigns a rendering backend per view (D3 for InfoVis, VTK.js for SciVis), resolves defaults, and generates the executable code and interactive controls. Per-view routing lets a single specification produce a linked, multi-view visualization. %spanning both InfoVis and SciVis.
This is the property \newsysname~exploits: every refinement yields a verifiable RaivenDSL artifact that can be checkpointed and recompiled rather than re-interpreted, making the specification, not the dialogue, the natural unit of version history.

\section{System Design}
\label{sec:system}
\begin{figure}
\centering
\includegraphics[width=1\columnwidth]{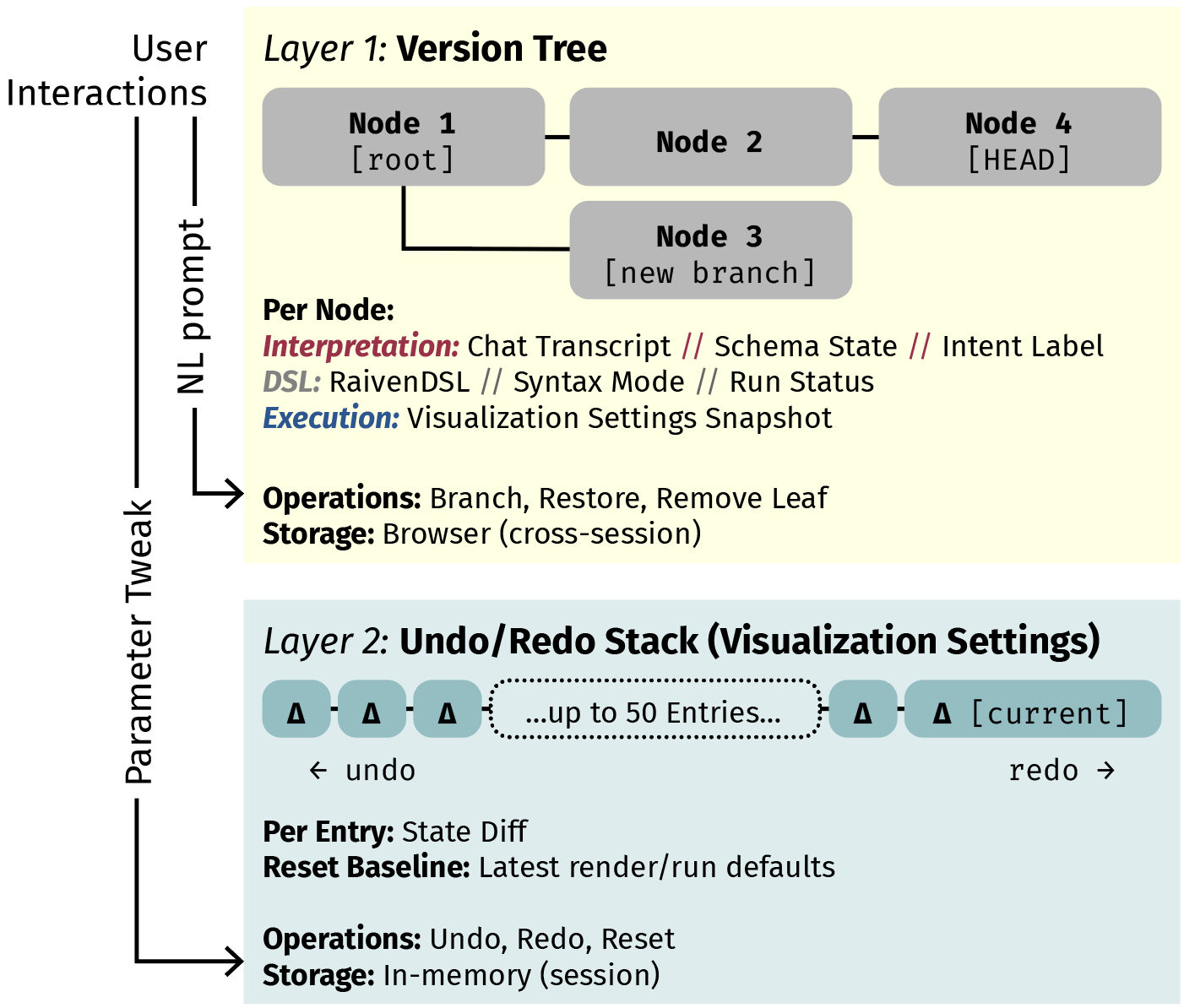}
\caption{\textbf{Two-level state management architecture.} Layer~1 persists a branchable version tree of DSL checkpoints across sessions; Layer~2 maintains a bounded undo/redo stack over visualization settings within the current rendered state. Navigating to a version node resets the stack baseline.}
\label{fig:system}
\end{figure}
\begin{figure}[t]
\centering
\includegraphics[width=0.87\columnwidth,trim={0 0 0 0},clip]{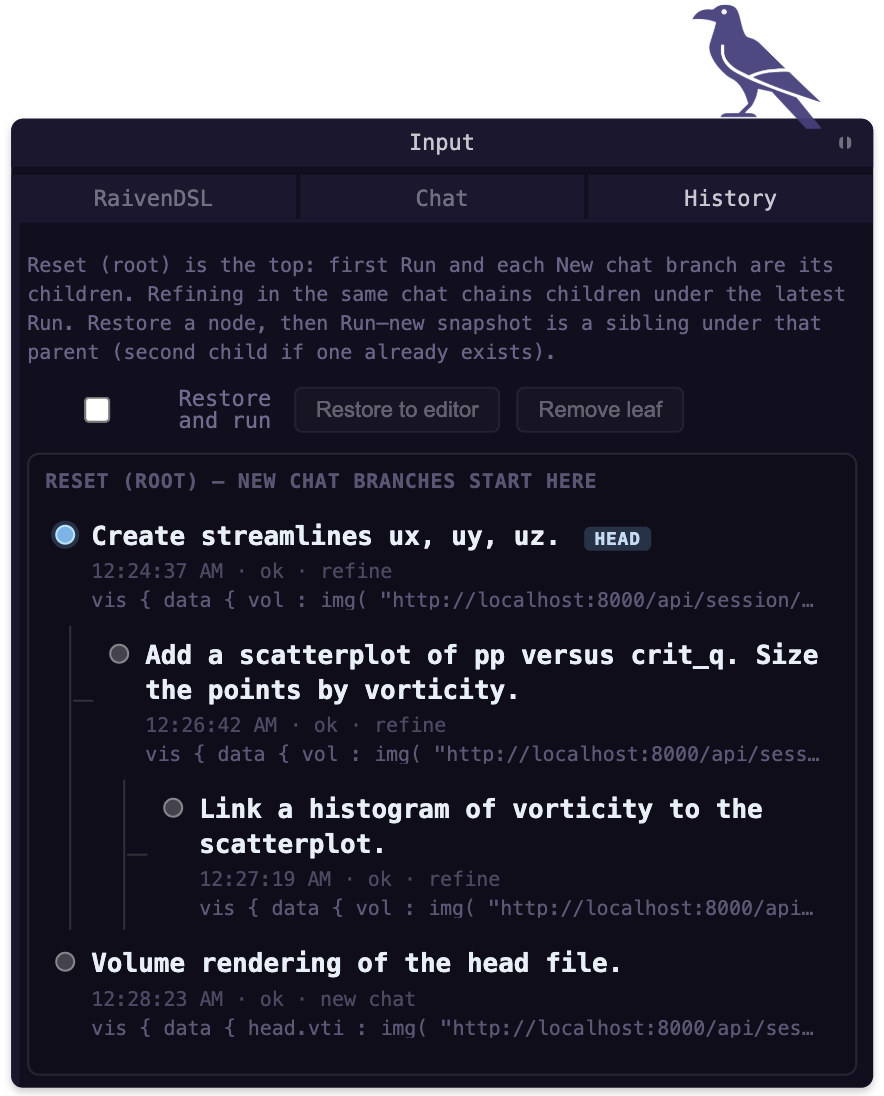}
\caption{%
\textbf{Version tree UI, shown on an illustrative example.} Each node is a RaivenDSL checkpoint, with indentation showing nesting and two different starting branches. The node text shows representative auto-generated labels: each summarizes the user intent behind that checkpoint, and a child's label describes its change relative to its parent. Visual marks distinguish the current HEAD and its ancestors.}
\label{fig:version-tree}
\end{figure}

\newsysname~extends the \sysname~pipeline with two complementary layers of state management: a persistent version tree over validated DSL checkpoints for coarse-grained restoration and branching, and an interaction-level undo/redo stack for fine-grained changes within the active rendered state. Both run client-side; the version tree persists in browser storage on a single client.
%so a session's history is confined to that browser and is not yet shared across devices or collaborators (Section~\ref{sec:discussion}).
%Alex: multi user discussed in discussion

\subsection{Version Tree}
\label{sec:versionGraph}
Each Run, the user action that compiles the current RaivenDSL specification and renders it, attempts to commit a snapshot into the version tree. If a Run's DSL is canonically identical to an existing node, the system reuses that node and updates HEAD rather than duplicating it. Otherwise it creates a new node capturing the DSL and session context at commit time: RaivenDSL text and syntax mode, run status, source lineage, a chat and schema panel snapshot, the latest user intent hint, and, typically on successful runs, persisted visualization settings. Nodes are stored in browser storage, persisting across reloads and sessions on the same client.

Version nodes are compact. RaivenDSL text is typically under 2KB; a full serialized node also stores the chat/schema snapshot and control state. The tree in Figure~\ref{fig:version-tree}, for example, serializes to roughly 30KB across its four nodes, each a few KB. Even trees far larger than any a user can navigate and retrace remain well within the browser's 5MB quota, so navigation usability, not storage, is what bounds tree scale in practice, consistent with participants flagging legibility well before any storage limit (Section~\ref{sec:designImplications}).

%Version nodes are compact. RaivenDSL text is typically under 2KB (median $\sim$0.4KB); a full serialized node, which also stores the chat/schema snapshot and control state, averages about 6KB, most of it the snapshot rather than the DSL. The tree is roughly 600KB for 100 nodes and well within the browser's 5MB quota. Backend rebuild on restore completes in under 10ms and recompilation in under 100ms; end-to-end latency is instead governed by dataset fetch and rendering, sub-second on a warm cache and a few seconds for first loads of remote volumetric data. Navigation usability, not storage or restore cost, is what bounds tree scale, consistent with participants flagging legibility well before any storage limit.

\subsection{Version History UI}
The version tree is exposed in a browsable history panel (Figure~\ref{fig:version-tree}). Users can select any node, inspect its DSL and run status, restore it to the editor and chat, or trigger a \emph{Restore and Run} that immediately recompiles the restored specification. Leaf nodes on abandoned paths can be removed. Visual marks distinguish the current HEAD and its ancestors, giving an at-a-glance map of the refinement history.

Node labels are generated automatically so the tree populates and stays maintained without the user naming each node by hand. The scheme is hybrid: a client-side title is derived from the user intent hint captured at commit time, giving every node an immediate label drawn from what the user asked for, and an LLM-generated summary from a backend labeling endpoint refines it. For a child node, this summary describes the change relative to its parent rather than restating the node's contents in isolation.
%since what distinguishes a node from its ancestor is usually more useful for navigation than a standalone description.
Deriving the base title client-side keeps labeling instant for every node, while the LLM summary adds this parent-relative detail where a terse intent hint alone would not disambiguate.

\subsection{State Restoration}
Selecting a node restores the complete session state across frontend and backend. The system replays the editor DSL and syntax mode, the chat transcript, and the schema panel, while a backend call rebuilds the agent's schema and conversation context and clones any schema-linked data files into a fresh session. Restoration is best-effort. The DSL and local UI restore first and unconditionally, so if backend rebuild or data cloning fails, the user is left with a usable local view and a warning rather than a blocked restore. 
%Because each checkpoint is a verifiable RaivenDSL specification rather than a raw dialogue transcript, restoration recompiles a fixed artifact rather than re-interpreting prior context, and the specification restored at a node is by construction the one checkpointed there.
Restoration recompiles the specification checkpointed at the node rather than re-interpreting prior context, and the restored specification is by construction the one committed there.
This reproducibility holds for the specification and control state but not for dataset contents, which are re-fetched at render time rather than snapshotted (Section~\ref{sec:discussion}). We treat specification-level reproducibility as an architectural property of versioning the DSL and leave its empirical validation to future work.

\subsection{Undo, Redo, and Reset}
Independently of the version tree, \newsysname~adds undo, redo, and reset to the visualization control panel. These operate on \emph{visualization-settings changes} within the current rendered visualization, not on DSL checkpoints. Changes are recorded as diffs, debounced at 300ms so that continuous slider drags collapse into a single undo step once the user pauses rather than recording every intermediate value, and kept in a bounded stack of up to 50 entries to cap client-side memory while retaining enough recent tweaks to recover from typical in-view adjustments. Longer-horizon workflow history is carried by the version tree, not undo/redo. Replaying a diff refreshes both UI bindings and renderer callbacks so the visualization updates immediately, and reset returns controls to the defaults established at the latest render/run.

The two layers are deliberately separate but complementary (Figure~\ref{fig:system}): version-tree branching gives coarse, semantic checkpoints for recovering from a wrong design direction or exploring an alternative encoding, while control-panel undo/redo gives fine-grained, low-latency rollback of parameter tweaks. Together they cover recovery at both the specification and parameter levels.

\section{User Study}
\label{sec:study}

\begin{figure*}[t]
  \centering
  \begin{minipage}[c]{0.71\textwidth}
    \centering
\includegraphics[width=\linewidth]{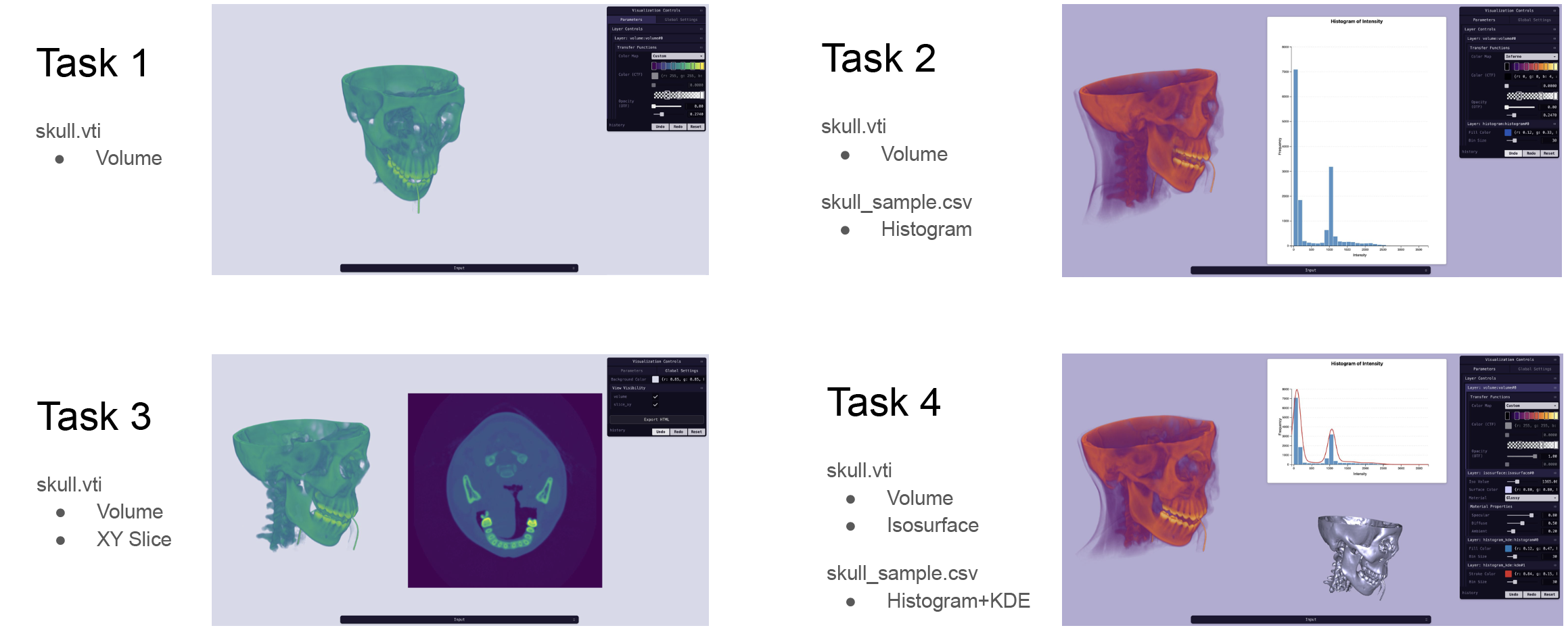}
  \end{minipage}%
  \hfill
  \begin{minipage}[c]{0.265\textwidth}
    \vspace{0pt}% align tops
    \small
    \textbf{Post-Session Interview Questions}
    \begin{enumerate}[leftmargin=*, itemsep=0pt, topsep=2pt]
      \item How useful did you find the version tree?
      \item Was there a moment where it saved you effort, or a moment
            where you wished it worked differently?
      \item Could you always tell where you were in the tree visually?
      \item Was it intuitive to use? If not, what was confusing and
            how could it be clearer?
      \item What would you change about how the versioning works?
      \item What kinds of workflows do you think this would be most
            useful for?
      \item I overall enjoyed the versioning history functionality. (1 = Disagree, 5 = Agree)
    \end{enumerate}
  \end{minipage}
  \caption{\textbf{Study tasks and post-session questions.} Each slide showed a target visualization, required files, and mark types.
  }
  \label{fig:tasks-and-questions}
\end{figure*}
To characterize how users adopt branchable conversational state during iterative refinement, we ran a formative pilot study with post-session interviews. It is exploratory by design: the aim is to observe how researchers engage with the version tree and to derive design directions, not to measure efficacy against a baseline. The results are early, promising evidence rather than confirmation; a controlled comparison against Raiven without version history, with measurable outcomes, is the planned next phase (Section~\ref{sec:discussion}). 

%We treat scientist-in-the-loop oversight as the motivating setting rather than the object of this study.

\subsection{Method}
Three visualization researchers (two PhD students and a postdoctoral researcher) each completed a single 20--30 minute session with \newsysname's versioning features.
%All tasks used a skull CT scan and a companion CSV of intensity values.
Each participant first watched a short video introducing the full feature set without prescribing when to use each feature.

Participants were then given four target screenshots to recreate, one at a time (Figure~\ref{fig:tasks-and-questions}).
%and were not told to work in any particular order,
They were not instructed
to revisit earlier states, or to branch; they chose how to reach each subsequent target using any combination of new prompts, restoration, and branching. The task sequence was designed so that a branching-aware path through it produces the tree shown in Figure~\ref{fig:teaser}: Task~1 yields the root (a volume rendering of the CT scan); Task~2 extends it linearly (volume + histogram from the CSV); Task~3's target (volume + XY slice) shares the root's volume rendering but not the histogram, so the efficient route branches from the root rather than continuing from node~2; and Task~4's target (volume + isosurface + histogram/KDE overlay) builds on the histogram, so its efficient route extends node~2. Whether participants took these routes or rebuilt states from scratch is part of what the study observes. Because participants authored each state through their own prompts, node labels reflect their wording rather than the task descriptions. The post-session interview covered tree utility, legibility, intuitiveness, desired changes, and anticipated use cases, plus a 5-point Likert rating of overall enjoyment (Figure~\ref{fig:tasks-and-questions}).

\subsection{Results}
\label{sec:results}
%We report what participants did, then what they reported in interviews, marking interpretation as such.
With three participants, these are qualitative observations, not quantitative findings. All three rated 5/5 on \textit{``I overall enjoyed the versioning history functionality''} (5-point Likert, 1 = disagree, 5 = agree).

\textbf{Branching behavior and tree adoption.} We imposed no structure on how participants reached each target, in order to see whether they would exploit the tree on their own. At Task~3, whose target shared the root's volume rendering but not the histogram, all three participants branched from the root, reusing the existing volume rather than rebuilding it. At Task~4, whose target builds on the histogram, two took the most efficient route and branched from node~2, where the volume and histogram were already in place and only the isosurface remained. The third branched from node~1 instead: this still reused the volume rather than starting over, but because node~1 lacked node~2's histogram, the participant regenerated the histogram before adding the isosurface and KDE overlay. We did not probe why they chose the higher ancestor; they may not have noticed node~2 was the more complete starting point, or may have preferred to rebuild the histogram themselves. Either way, all three exploited the tree to reuse prior work rather than restart, and the only divergence was how far up the tree one participant branched. With the caveat that the task sequence made branching a natural route, this suggests participants readily adopted the tree as a working mechanism, while the Task~4 divergence shows the most efficient branch point is not always the one chosen absent guidance.

\textbf{Perceived value and mental model.} Participants articulated the tree's value primarily as avoiding rework: two independently noted that restoring and branching let them continue from a prior state rather than rebuilding from scratch. The Task~4 divergence is a mild counterpoint: branching from node~1 rather than node~2 still reused prior work, but less of it than the tree made available. All could locate themselves in the tree throughout the session, though one qualified this for larger trees: \textit{``yes if I only have a small number of branches or leaves, however, there will be more nodes later and I will not find my current position.''} Node labels were the primary navigation aid, and one participant cited seeing the generating prompt at each node as key to orientation. For larger trees, participants proposed a git-style tree panel in the chat interface and per-node thumbnails of the rendered visualization for rapid differentiation across branches.

\textbf{Undo/redo vs.\ restore.} No participant confused undo/redo with version-tree restore; each reached for one or the other unprompted, using the tree to redirect a design and undo/redo to correct a parameter. This suggests the granularity distinction between the layers is perceivable without instruction, though the evidence is thin: only one participant used undo/redo at all, to fix a mistaken color-theme change, while another found the control panel direct enough that the task did not call for it. Because the tasks required no sustained parameter tuning, this uptake reflects task design as much as the interface.

\textbf{Anticipated use cases.} Participants saw the tree as most useful for long-horizon, multi-session work, including iterative figure design for papers, collaborative exploration, and domain analyses such as gene expression, settings where analysis unfolds over many steps and alternatives must be revisited, none tested here. Two participants noted the study tasks were too small to exercise the tree fully, wanting more challenging or longer cases to reveal its benefits and failure modes.

\subsection{Design Directions}
\label{sec:designImplications}

The pilot surfaces three design directions. Each follows from the observations above but points to work this study does not carry out.

\textbf{Thumbnails for visual orientation.} A participant asked for per-node thumbnails of the rendered output, so that branches could be told apart by appearance rather than label text alone. This matches a need likely to grow with tree size: where sibling states differ in ways a short label does not capture, a rendered preview offers a fast visual check before a restore. The Task~4 divergence is suggestive here, since the two candidate nodes differed by only a single clause, a thumbnail
would
have made the more complete starting point visible at a glance. Whether thumbnails improve navigation in practice remains to be tested.

\textbf{Labels carry the navigation load.} Participants oriented themselves primarily through node labels, so label distinctiveness bounds navigation accuracy as trees grow. \newsysname~already generates parent-relative labels, but as trees deepen and more nodes share structure, short labels alone are unlikely to remain sufficient. This points less toward richer label text than toward complementary cues, such as thumbnails or label search, that disambiguate where wording cannot.

\textbf{Version trees must scale.} Participants navigated confidently at study scale (four nodes) but anticipated difficulty as branches accumulate. Tree pruning, collapsed linear chains, label search, and sub-branch summaries are candidate responses, but the pilot neither exercises trees at scale nor evaluates these mechanisms; scalability is a problem the study identifies rather than solves.

\section{Discussion \& Future Work}
\label{sec:discussion}

\newsysname~shows that a verifiable DSL is a practical foundation for versioned conversational visualization, supporting iterative analysis rather than only first-pass generation. Grounding history in the specification (Section~\ref{sec:versionGraph}) also changes what versioning is for. Classical history lets users revisit states they authored themselves; oversight of an AI-driven workflow instead means inspecting states an agent produced, recovering decisions the human did not directly make. Because each node exposes the specification the agent committed to, that decision is independently checkable, which a black-box pipeline or a dialogue transcript cannot offer. Branchable DSL history is thus not just a convenience for the author but a substrate for oversight.

Steering is its active counterpart. When an agent's trajectory goes wrong, restoration provides the vantage point and branching the redirection: instead of correcting through further prompts, the user returns to an earlier verified state and opens a new path from it. The two-level design separates this by grain, with branching as coarse steering at the specification level and undo/redo as fine steering within the rendered state. We frame both as affordances the architecture provides, not behaviors this pilot measures. Undo/redo uptake was constrained by task design rather than discoverability (Section~\ref{sec:results}); future studies should include parameter-tuning tasks, such as matching a target transfer function, to test adoption where it is needed.

\textbf{Limitations.} The central limitation is scope: three visualization researchers, four short branching-oriented tasks, a single dataset, no baseline. Because the tasks were structured to elicit branching with no control condition (linear history only, or no versioning), observed behaviors reflect adoption under conditions that favored it and cannot be causally attributed to branching. The short, single-dataset sessions exercise only small trees (four nodes) and do not test long-horizon or multi-session use. Each task did compose views across both backends in a single specification (a volume rendering on VTK.js with a histogram or KDE overlay on D3), so the version tree checkpointed and restored multi-backend specifications; genuine cross-backend operation, beyond composing backends within one spec, remains future work in Raiven itself. Participants were visualization researchers,
%not the domain scientists the system targets,
so whether the same patterns hold for non-visualization-expert users remains open. The reproducibility of restore is also narrower than it may sound: version nodes store the RaivenDSL specification, control state, and schema metadata but not dataset contents, so re-fetched data that has been deleted or modified can cause load failures or silently yield a visualization from current rather than commit-time state; full point-in-time reproducibility would require content-addressed data snapshots. Finally, the version tree lives in browser storage on a single client (Section~\ref{sec:versionGraph}), so history is confined to that client, and cross-device or collaborative sharing is not yet supported.

\label{sec:conclusion}
\textbf{Future Work.} The most immediate next step is the controlled evaluation this pilot lacks: comparing Raiven with and without the version tree on tasks that include genuine wrong turns, measuring task-completion time, recovery success, and rework avoided, so that benefit can be attributed to branching rather than inferred. Beyond this, a clear direction is to study how the tree behaves when refinements come from agents rather than hand-authored, creating histories that may grow deeper, faster, and less predictably, and whether the branching and undo/redo affordances support oversight and steering at that scale. Such histories will also call for capabilities the current system lacks, including node annotations, branch merging, history search, and cross-session sharing, which prior provenance research suggests become important at scale~\cite{cutler2020trrack,xu2020survey,dunne2012graphtrail} and would support teams on shared, evolving workflows.

\section*{Acknowledgments}
This work was supported by the U.S. Department of Energy, Office of Science, Office of Advanced Scientific Computing Research's Computer Science Competitive Portfolios program under Contract No. DE-AC05-00OR22725. This work was partially supported by NIH grant 1U01CA284207.

\bibliographystyle{abbrv-doi-hyperref}
\bibliography{bibliography}

\appendix
\onecolumn
\section{Study Task Instructions}
\label{appendix:tasks}

Participants received the following slides during the study session,
each showing a target visualization screenshot, the required files
and the respective mark types. Participants were not told the names of any system features.

\bigskip
\subsection{Task 1}
%\textbf{Files:} \texttt{skull.vti} \quad \textbf{Marks:} Volume
\begin{figure}[h]
  \centering
  \fbox{\includegraphics[width=1.0\linewidth]{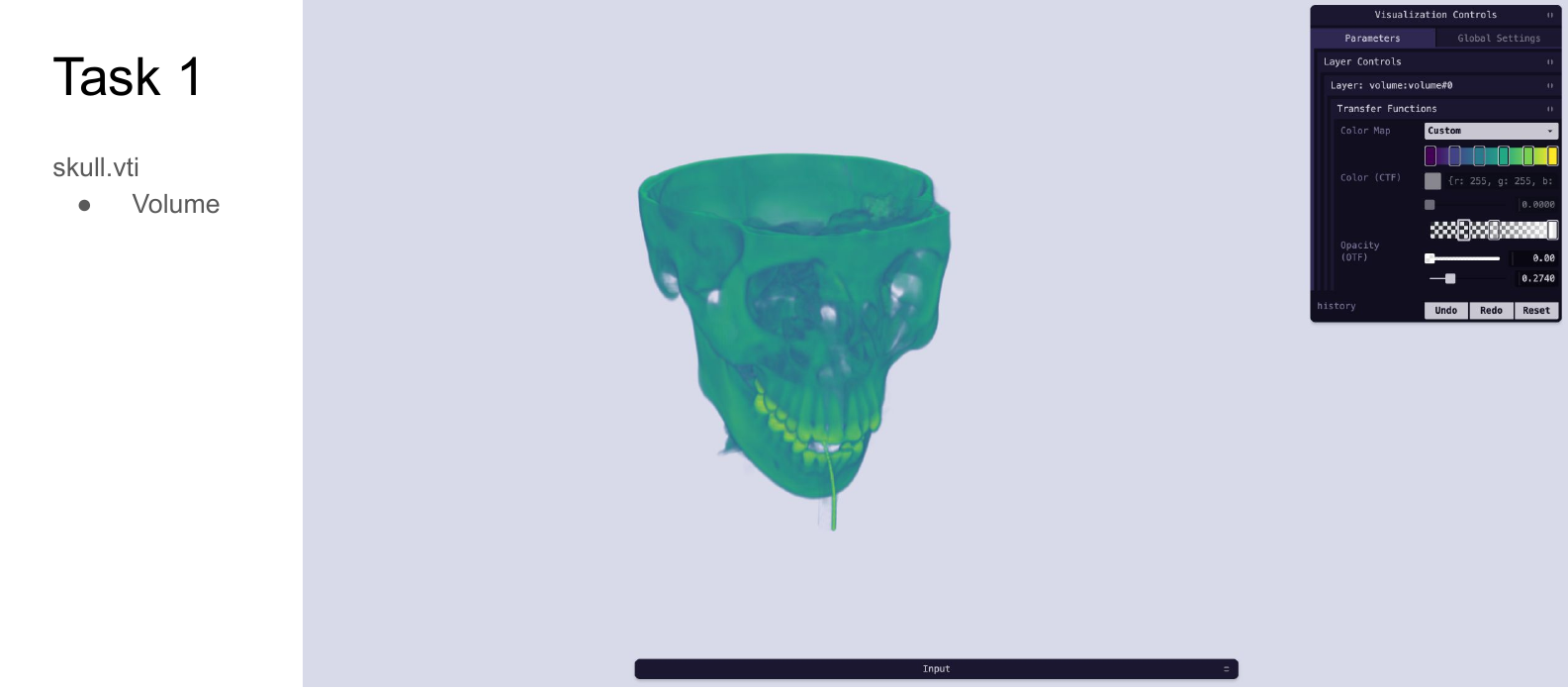}}
\end{figure}

\subsection{Task 2}
%\textbf{Files:} \texttt{skull.vti}, \texttt{skull\_sample.csv} \quad \textbf{Marks:} Volume, Histogram
\begin{figure}[h]
  \centering
  \fbox{\includegraphics[width=1.0\linewidth]{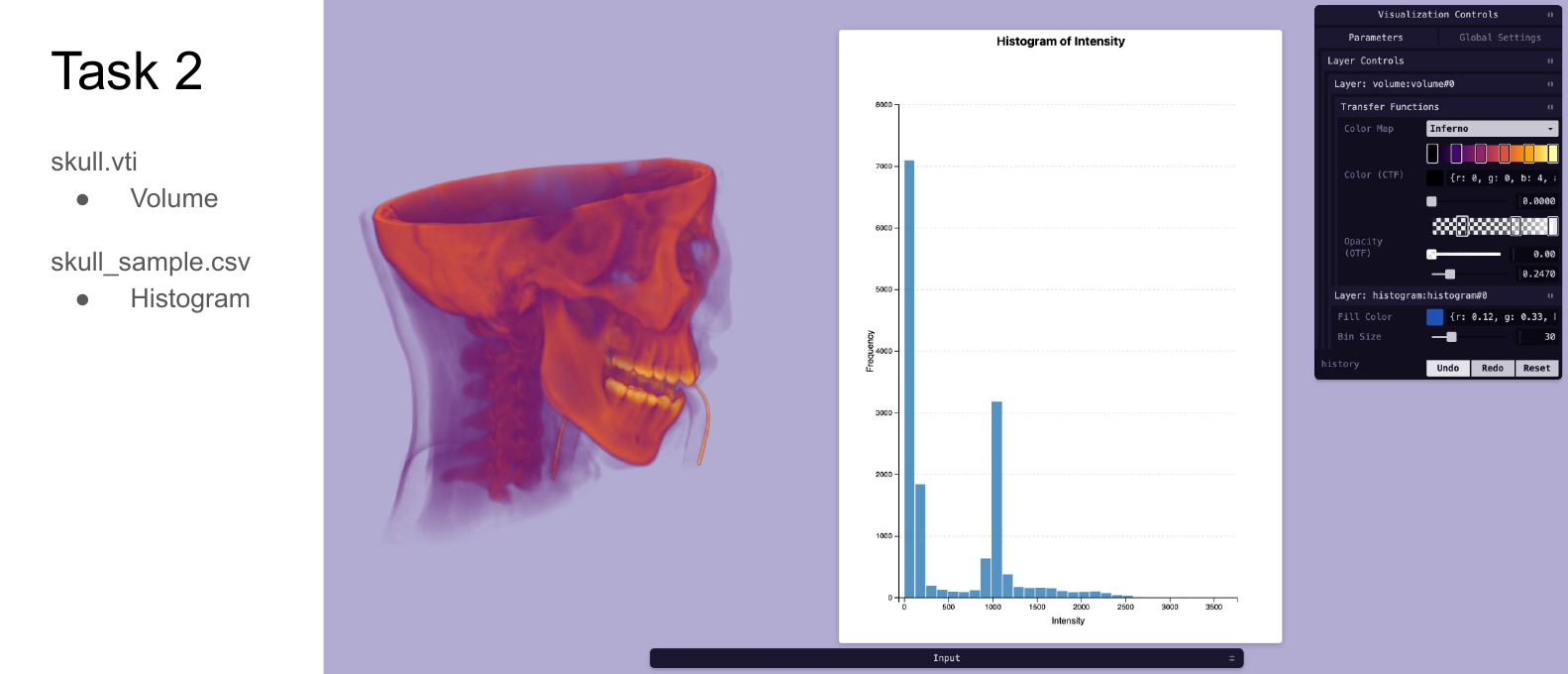}}
\end{figure}

\clearpage

\subsection{Task 3}
%\textbf{Files:} \texttt{skull.vti} \quad \textbf{Marks:} Volume, XY Slice
\begin{figure}[h]
  \centering
  \fbox{\includegraphics[width=1.0\linewidth]{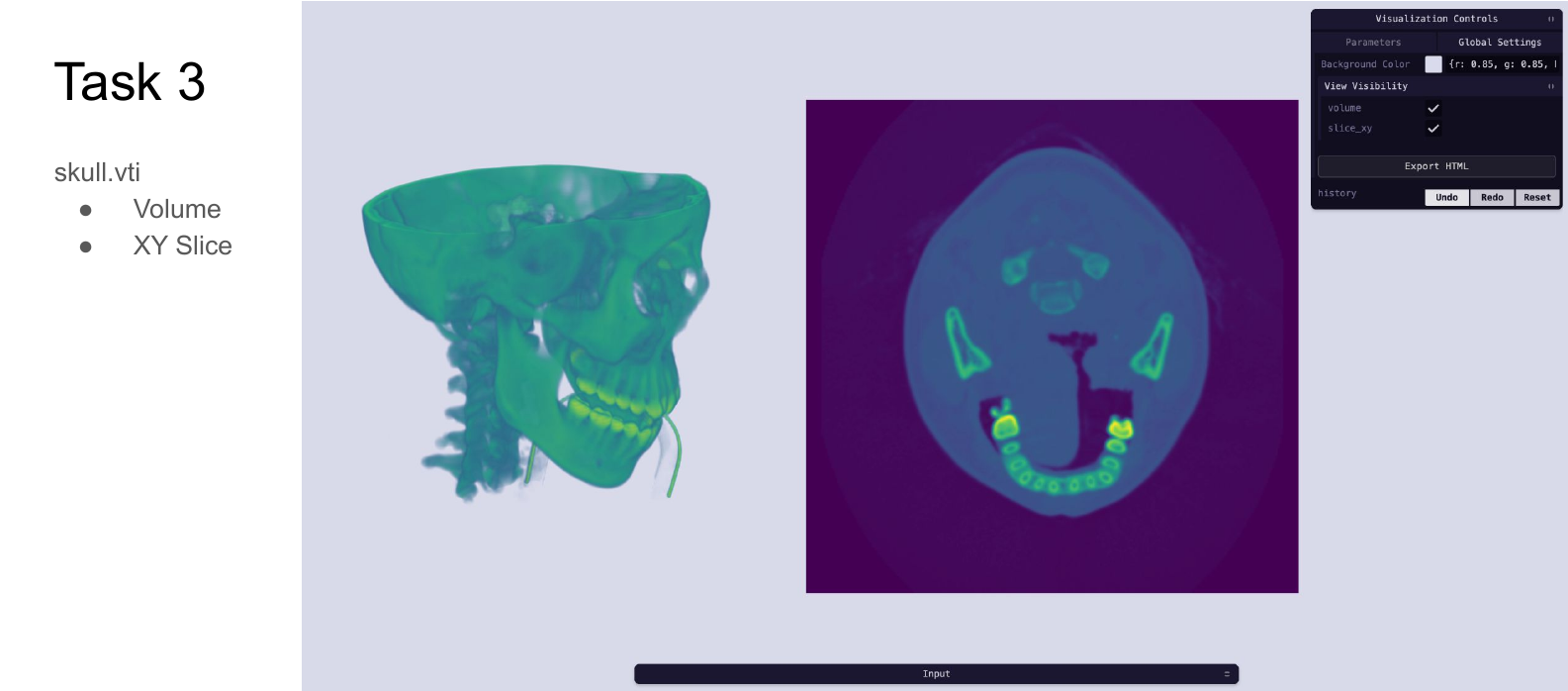}}
\end{figure}

\subsection{Task 4}
%\textbf{Files:} \texttt{skull.vti}, \texttt{skull\_sample.csv} \quad \textbf{Marks:} Volume, Isosurface, Histogram + KDE
\begin{figure}[h]
  \centering
  \fbox{\includegraphics[width=1.0\linewidth]{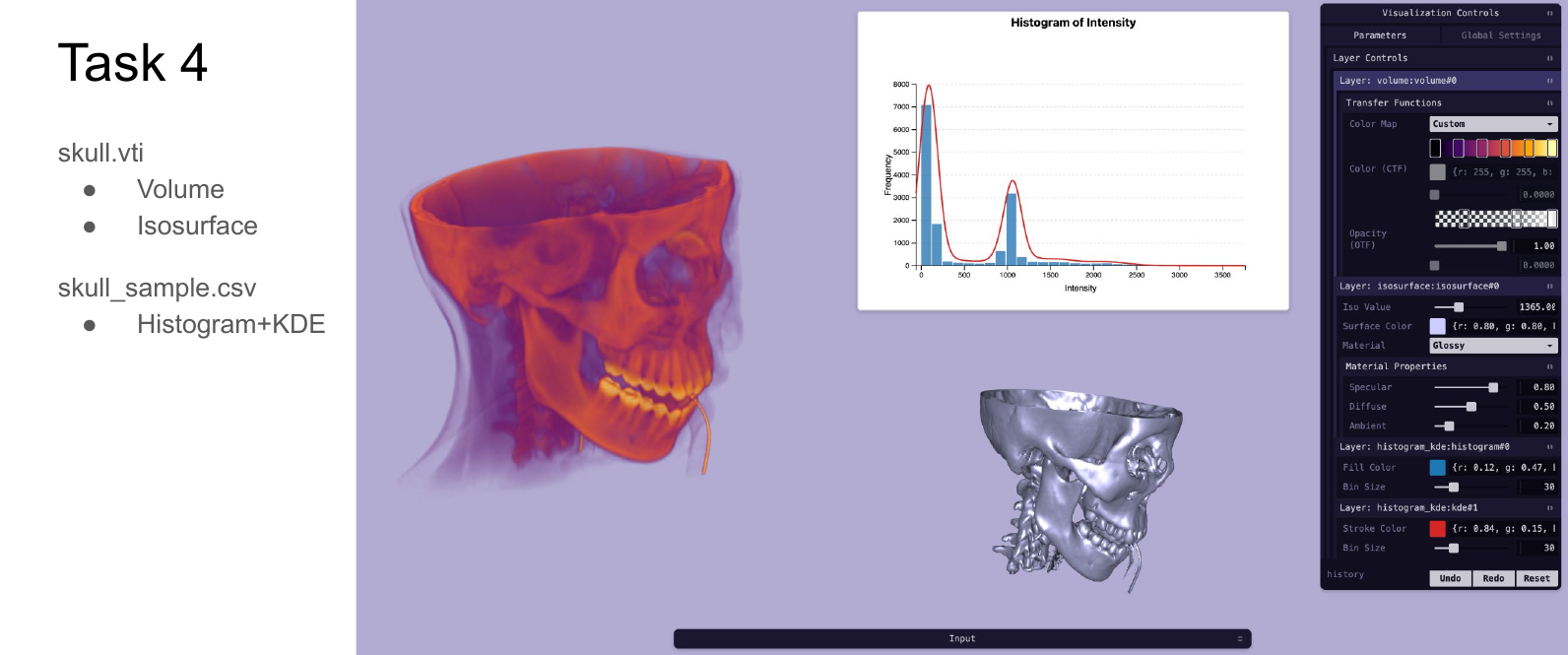}}
\end{figure}

\section{Post-Session Interview Questions}
\label{appendix:interview}

\begin{enumerate}
  \item How useful did you find the version tree?
  \item Was there a moment where it saved you effort, or a moment
        where you wished it worked differently?
  \item Could you always tell where you were in the tree visually?
  \item Was it intuitive to use? If not, what was confusing and
        how could it be clearer?
  \item What would you change about how the versioning works?
  \item What kinds of workflows do you think this would be most
        useful for?
\end{enumerate}

\end{document}